\documentclass[prd,aps,amsmath,superscriptaddress,twocolumn,amssymb,nofootinbib]{revtex4}
\usepackage{amsmath,amssymb,hyperref,graphics,epsfig,subfigure}
\usepackage{color}

\begin{document}
\newcommand {\nn}    {\nonumber}
\renewcommand{\baselinestretch}{1.3}
\title{Thick Brane Split Caused by Spacetime Torsion }
\author{ Jie Yang\footnote{yangjiev@lzu.edu.cn, corresponding author},
         Yun-Liang Li\footnote{liyunl09@lzu.edu.cn},
         Yuan Zhong\footnote{zhongy2009@lzu.edu.cn},
         Yang Li\footnote{liyang09@lzu.edu.cn} }
\affiliation{Institute of Theoretical Physics,
              Lanzhou University, Lanzhou 730000,
             People's Republic of China}

\begin{abstract}
In this paper we apply the five-dimensional $f(T)$ gravity with $f(T)=T+k T^n$
to brane scenario to explore the solutions under a given warp factor, and we find that the analytic domain wall solution will be a double-kink solution when the geometric effect of spacetime torsion is strongly enhanced. We also investigate
the localization of fermion fields on the split branes
corresponding to the double-kink solution.
\end{abstract}




\maketitle

\section{Introduction}

The presence of extra dimensions is playing a fundamental role in
solving the hierarchy problem, explaining physical interactions
based on common principles and other problems in high energy physics
\cite{Randall1999,Karch2001b,Goldberger1999,Garriga2000a,Gremm2000a,Csaki2000a,Csaki2000,Csaki2004,DeWolfe2000a}.
Under the condition of no undesirable physical consequences
obtained, as we know so far, any realistic candidate for a
 grand unified theory should be multidimensional. Because of the absence of observational and experimental data, preference makes no difference in discriminating
 various kinds of multidimensional models of gravity. Actually, all sorts of models have been studied in extra dimension gravity.

The concept of brane scenario was introduced  in 1983 by Robakov and Shaposhnikov, who pointed out that we live in a topological defect embedded in 5-dimensional spacetime, i.e., domain wall, or thick brane in modern terminology which was used as a new approach to solve the problem of the unobservability of the extra dimensions \cite{Rubakov1983bb}. According to the idea, particles corresponding to electromagnetic, weak and strong interactions are confined on some hypersurface called a brane. Only gravitation and some exotic matter could propagate in the extra dimension. And in \cite{Rubakov1983bb} the authors found that particles with spin 0 and 1/2 can be trapped on the domain wall described by a scalar field without gravity. During the 80s and the early 90s, one of the most striking facts which activated the studies on brane models was the development in superstring theory and M-theory since the mid of 90s, especially the discovery of D-brane solutions \cite{Polchinski1995mt,Polchinski1996fm}. In 1999, Randall and Sundrum (RS) proved that gravitation also can be localized on the brane if one takes the gravity into consideration \cite{Randall1999}. This is the famous RS brane model which attracts much concentration from physicists because of its theoretic value, observable effect and solving the long-standing hierarchy problem and cosmological constant problem. And graviton resonances were previously considered in thick brane scenario in \cite{Gremm2000a,Csaki2000a}.

So far, various thick brane or domain wall solutions have been investigated (a review in Ref. \cite{Dzhunushaliev2009va}) and the trapping of all kinds of matter fields on the single-brane or multi-branes are also discussed for both thin or thick branes \cite{RandjbarDaemi2000ft,RandjbarDaemi2000cr,Kakushadze2000zp,Youm2000dc,Oda2000dd,Bajc2000,Oda2000,Arias2002,Barbosa-Cendejas2005,
Barbosa-Cendejas2006,Barbosa-Cendejas2008,Liu2008,Liu2008c,Liu2008a,Liu2009c,Liu2009dw,Bazeia2009,HerreraAguilar2010kt}. All of these works only considered the contribution of spacetime curvature without torsion. In this paper we would like to investigate thick brane solutions caused by the spacetime torsion. An applicable theory is the teleparallel equivalent of General Relativity (TEGR)\cite{Hayashi1979,Aldrovandi2007,AndradePereira1997,Andrade2000,
AndradeGuillenPereira2000,Sousa2010a} which instead of using the curvature defined via the Levi- Civita connection, it uses the Weitzenb\"{o}ck connection that has no curvature but only torsion. This theory allows us to interpret general relativity as a gauge theory for a translation group. And in this context, gravity is not due to curvature, but to torsion, and torsion accounts for gravitation not by geometrizing the interaction, but by acting as a force.

A question that will be asked is that what is the role of torsion or the difference between torsion and curvature \cite{Hayashi1979,Arcos:2005ec}. Although the equations of motion in teleparallel gravity are dynamically equivalent to those in general relativity and relate to the same degrees of freedom of gravity (more general relativity theories, like Einstein-Cartan and gauge theories for the Poincar\'{e} and the affine groups, consider curvature and torsion as representing independent degrees of freedom), the teleparallel gravity describes a different geometry, the Weitzenb\"{o}ck spacetime. The spacetime metric $g_{\mu\nu}$ plays no dynamical role in the teleparallel description of gravitation.

If we want to investigate the influence of spacetime torsion, we should modify the teleparallel gravity. Following the spirit of $f(R)$ gravity (see \cite{Sotiriou2010} for a review, \cite{Parry2005,Afonso2007,Bouhmadi-Lopez2010,Dzhunushaliev2010,Zhong2010ae} for applications in braneword), a generalization of teleparallel gravity is $f(T)$ gravity which was first proposed by Bengochea and Ferraro to explain the observed acceleration of the universe \cite{Bengochea:2008gz}. And models based on modified teleparallel gravity were also found to provide an alternative to inflation without inflaton \cite{Ferraro:2006jd,Ferraro:2008ey}. It therefore has attracted some attention recently. More recently, Linder \cite{Linder2010} proposed two new $f(T)$ models to explain the accelerating expansion and found that the $f(T)$ theory can unify a number of interesting extensions of gravity beyond general relativity.

The fact that we should note is that $f(T)$ gravity could be a phenomenological extension of the teleparallel gravity, inspired by the $f(R)$ generalization of the general relativity. Although the $f(R)$ gravity is probably not the low-energy limit of some fundamental theory, it does include models that can be motivated by effective field theory. In contrast, $f(T)$ gravity seems at this stage to be just an \textit{ad hoc} generalization. Recently it was pointed out that $f(T)$ gravity violates the local Lorentz invariance \cite{Sotiriou2010mv,LiSotiriouBarrow2011}. Nevertheless, it still attracts an increasing interest in the literature because of its advantage over $f(R)$ gravity, namely, its field equations are at most second-order instead of fourth-order.
The validity of $f(T)$ gravity as an alternative also has been investigated by analyzing the large-scale structure \cite{LiThomasSotiriou2011} and the observational constraints on model parameters \cite{Wu2010a,Bengochea:2010sg}. Apart from obtaining acceleration, one can reconstruct a variety of cosmological evolutions \cite{Myrzakulov:2010vz,Myrzakulov:2010tc,Yerzhanov:2010vu,Wu:2010xk},  can consider the possibility of the phantom divide crossing \cite{Wu:2010av,HamaniDaouda2011,Karami:2010bu,Bamba:2010iw}, and can investigate the vacuum and matter perturbations \cite{Chen:2010va,Zheng:2010am,Dent:2011zz} beyond the background evolution. For other investigations, see for examples \cite{Miao1,Miao2}.

Considering the increasing interest in $f(T)$ gravity and the possibility as an alternative to general relativity, in this paper we investigate the impact of torsion instead of curvature on the structure of thick branes and the localization of fermions on the thick branes. The spacetime torsion can result in the splitting of the thick brane with an internal structure which is called double-kink defect since it seems to be composed of two standard kinks. In the works of Bazeia and his collaborators \cite{Bazeia:2002xg,Bazeia:2005tm,Afonso:2007mr,BazeiaMenezesMenezes2003,BazeiaFurtadoGomes2004} a class of defect structures were obtained by a $\phi^4$ potential. The appearance of the structure will result in a split in the matter energy density in the center of the brane. And the resonances of gravitons and fermions in the such structure scenario have been considered in recent works \cite{Almeida:2009jc,Cruz:2011kj,Cruz:2011ru,Chumbes:2010xg,Liu:2009ve,ZhaoLiu2011JHEP}. In our work we use the resonance detecting method to analyze the KK modes of fermion fields and how the internal structure due to geometric effect influence the resonances of fermion in the splitting brane.

The paper is organized as follows: In Sec. \ref{section2}, we first give a brief review of the teleparallel gravity and then give the field equations for the five-dimensional $f(T)$ brane. In Sec. \ref{section3}, because their equations are still second order, we obtain some exact analytic domain wall solutions for a given warped factor. In Sec. \ref{section4}, we study the localization of fermion fields on the thick branes by presenting the potential of the Schr\"odinger equations.

\section{Set ups and dynamical equations}
\label{section2}
Before we set up our model, let us briefly give a review of the teleparallel gravity. In teleparallel gravity, it is the vierbein or tetrad fields, $h_{a}(x^\mu)$ (rather than the metric) that work as the dynamical variables. At each point of the manifold, the tetrad fields form an orthonormal basis for the corresponding tangent space of the point. In four-dimensional teleparallel gravity, Latin indices $a,b,\ldots$ and Greek indices $\mu,\nu,\ldots$ both run from 0 to 3, label coordinates of the tangent space
and the spacetime, respectively. For a specified spacetime coordinate basis the components of $h_{a}(x^\mu)$ are $h^{\mu}_a$. Clearly, $h^{\mu}_a$ are both spacetime vectors
and Lorentz vectors.

The relation between the tetrad fields and the metric is given by
\begin{equation}\label{1}
    g_{\mu\nu}=\eta_{ab}h^{a}_{\mu}h^{b}_{\nu},
\end{equation}
where $\eta_{ab}=\text{diag}(-1,1,1,1)$ is the Minkowski metric for the
tangent space. From the relation~(\ref{1}), it follows that
\begin{equation}
    h^{\mu}_{a}h^{a}_{\nu}=\delta^{\mu}_{\nu}, \qquad h^{\mu}_{a}h^{b}_{\mu}=\delta^{b}_{a}.
\end{equation}

Instead of using the Levi-Civita connection $\Gamma^{\rho}_{~\mu\nu}$, we would like to apply the Weitzenb\"{o}ck tensor
\begin{equation}
    \tilde{\Gamma}^{\rho}_{~\mu\nu}=h^{\rho}_{a}\partial_\nu h^{a}_{\mu},
\end{equation}
and the torsion
\begin{equation}
   T ^{\rho}_{~\mu\nu}=\tilde{\Gamma}^{\rho}_{~\nu\mu}-\tilde{\Gamma}^{\rho}_{~\mu\nu},
\end{equation}
to establish the teleparallel gravity.
The difference between the Levi-Civita connection and Weitzenb\"{o}ck
connection is the well-known contortion
tensor~\cite{Weitzenbock1923}
\begin{equation}
  K^{\rho}_{~\mu\nu}\equiv \tilde{\Gamma}^{\rho}_{~\mu\nu}-\Gamma^{\rho}_{~\mu\nu}=\frac{1}{2}[T^{~\rho}_{\mu~\nu}+T_{\nu~\mu}^{~\rho}-T^{\rho}_{~\mu\nu}].
\end{equation}
By defining a tensor $S_{\rho}^{~\mu\nu}$:
\begin{equation}
S_{\rho}^{~\mu\nu}=\frac{1}{2}[{K^{\mu\nu}}_{\rho}
   -{\delta^{\nu}_{\rho}{T^{\lambda\mu}}_{\lambda}}
   +{\delta^{\mu}_{\rho}{T^{\lambda\nu}}_{\lambda}]},
\end{equation}
one can write the Lagrangian of the teleparallel gravity as
\cite{AndradePereira1997,Andrade2000,AndradeGuillenPereira2000,Hayashi1979,Aldrovandi2007}
\begin{eqnarray}
\label{Lagrangian}
L_T&=&-\frac{c^4h}{16\pi G}T=-\frac{c^4h}{16\pi G}S_{\rho}^{~~\mu\nu}T^{\rho}_{~~\mu\nu}\nonumber\\
&=&-\frac{c^4h}{16\pi G}
  \Big[ \frac{1}{4}T^{\rho}_{~~\mu\nu}T^{~\mu\nu}_{\rho}
       +\frac{1}{2}T^{\rho}_{~~\mu\nu}T_{~~~\rho}^{\nu\mu}
       -T^{~~~\rho}_{\rho\mu}T_{~~~\nu}^{\nu\mu}
  \Big],\nonumber\\
\end{eqnarray}
where $h=\det(h^{a}_{\mu})=\sqrt{-g}$, with $g$ the determinant of the metric $g_{\mu\nu}$. It is well known that the teleparallel gravity is equivalent to general relativity. Therefore, in order to discuss the effects of the torsion, we have to generalize the gravity.

As to the $f(T)$ gravity, we need only to replace the $T$ in Lagrangian \eqref{Lagrangian} by an arbitrary differentiable function of $T$, and then the action in five-dimensional
gravity is
\begin{equation}\label{action}
  S=-\frac{1}{4}\int d^5x h f(T)+\int d^5x \mathcal{L}_M,
\end{equation}
where we have taken $\frac{c^4}{4\pi G_5}=1$ for convenience.
The corresponding field equations read
\begin{equation}\label{fieldequation}
 h^{-1}f_T\partial_{Q}(h S_{N}^{~~MQ})+f_{TT}S_{N}^{~~MQ}\partial_Q T-t_{N}^{~~M}=-T_{N}^{~~M},
\end{equation}
where $f\equiv f(T), f_T\equiv\partial f(T)/\partial T, f_{TT}\equiv\partial^2 f(T)/\partial T^2$ and
 $t_{N}^{~M}=f_T\tilde{\Gamma}^{R}_{~SN}S_{R}^{~MS}-\frac{1}{4}\delta_{N}^{~M}f$, $T_{N}^{~M}$ is
  the energy-momentum tensor of the matter field. Capital Latin indices $M,N\ldots=0,1,2,3,5$. Here the field equations are expressed in purely spacetime form, not containing coordinates of the tangent space.

In our work we consider the static flat braneworld scenario with the metric
\begin{equation}\label{metric}
   ds^2=e^{2A(y)}\eta_{\mu\nu}dx^\mu dx^\nu+dy^2,
\end{equation}
where $\eta_{\mu\nu}=\text{diag}(-1,1,1,1)$ is the four-dimensional Minkowski metric, and $e^{2A(y)}$ is the warped factor.
Then the tetrad fields are $h^{a}_{\mu}=\text{diag}(e^{A},e^{A},e^{A},e^{A},1)$, $T=-12A'^2$.
From now on, the prime always denotes the derivative with respect to $y$, unless specified.

In our model, we take $f(T)=T+k T^n$, and $\mathcal{L}_M=h(-\frac{1}{2}\partial^M\phi~\partial_M\phi-V(\phi))$ \cite{Aldrovandi2007},
  where $\phi\equiv\phi(y)$ depends only on the extra dimension $y$.
And then the field equations are given as follows
\begin{eqnarray}
  &&~~~~~{\phi''}+4A'\phi'=\frac{d V(\phi)}{d\phi},
  \end{eqnarray}
\begin{eqnarray}
   &&\frac{1}{4}\left[12A'^2+(-1)^{n-1}12^nk(2n-1)A'^{2n}\right]\nonumber\\
  &&~~~~~~~~~~~~~~~~~~~~~~~~~~~~~~~~~~=-V+\frac{1}{2}\phi'^2,\label{equ2}\\
  &&(-1)^{n-1}2^{2n-3}3^nk(2n-1)A'^{2n-2}(2A'^2+n A'')\nonumber\\
 &&~~~~~~~~~~~~~~~+3A'^2+\frac{3}{2}A''=-V-\frac{1}{2}\phi'^2\label{equ1}.
\end{eqnarray}
Note that there are only two independent equations in the above equations.
Therefore, we need only to consider eq.~(\ref{equ2}) and the following one:
\begin{equation}
\label{main}
    \phi'^2=-\frac{[12A'^2+(-1)^{n-1}12^nkn(2n-1){A'^{2n}}]A''}{8A'^2},
\end{equation}
which is obtained by the combining of eqs.~(\ref{equ2}) and (\ref{equ1}).

\section{Solutions for $f(T)$ brane}\label{section3}

Although the model is a second-order derivative theory,
it is hard to give an analytic solution for general cases.
For simplicity, let us take
\begin{equation}
\label{condition}
   e^{2A(y)}=\cosh^{-2b}(\alpha y),\quad(b>0),
\end{equation} and consider the following cases.

\subsection{$n=\frac{1}{2}$}

For $n=1/2$, eqs.~(\ref{equ2}) and (\ref{equ1}) reduce to
\begin{eqnarray}
&&3A'^2= -(V-\frac{1}{2}\phi'^2),\\
&&3A'^2+\frac{3}{2}A''= -(V+\frac{1}{2}\phi'^2).
\end{eqnarray}
They are the same equations as those in Refs.~\cite{Bazeia2009,Gremm2000a}, where the gravity is described by general relativity. As a consequence,
the solutions of this case are equivalent to those in the case $f(T)=T$.
A domain wall solution has been obtained in~\cite{Bazeia2009,Gremm2000a}
by using a superpotential approach:
\begin{eqnarray}
    &&\phi(y)=\sqrt{6b} \arctan(\tanh(\frac{\alpha y}{2})),\\
    &&V(\phi)=\frac{3b\alpha^2}{4}\Big[(1+4b)\cos^2(\frac{2\phi}{\sqrt{6b}})-4b\Big].
\end{eqnarray}
Obviously, $\alpha$ is a parameter which fixes the thickness of the wall.
As stated in Ref. \cite{Gremm2000a}, as $y\to\pm\infty$, $A(y)\to{-b\alpha|y|}$. Thus, the spacetime described by the metric (\ref{metric}) and \eqref{condition} is asymptotically
$AdS_5$.

\subsection{Other positive integers $n$}

With the substitution of \eqref{condition} into \eqref{main}, we should have the following condition
\begin{equation}\label{constraint}
\Delta\equiv(-1)^n12^{n-1}n(2n-1)kb^{2n-2}\alpha^{2n-2}\leq1,
\end{equation}
to make the right side of \eqref{main} non-minus.
Only when \eqref{constraint} is satisfied, we can obtain a real function solution. Note that the existence of $(-1)^n$ constraints the values of $n$ and $k$.

\begin{figure}[h]
\begin{center}
\includegraphics[]{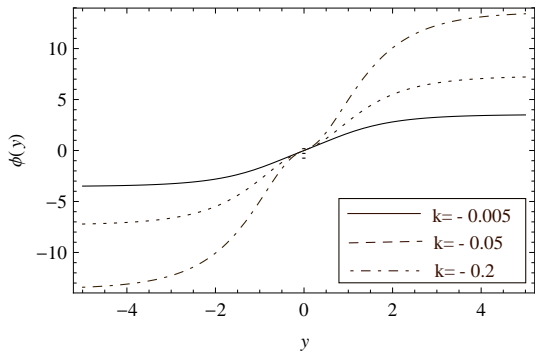}
\end{center}
 \caption{The shape of the scalar $\phi(y)$ plotted with $n=2, b=1, \alpha=1$. With the decrease of $k$, double-kink solutions will be more notable.}\label{phi}
\end{figure}

For $n=2$, we yield an analytic domain wall solution:
\begin{eqnarray}\label{solution2}
 &&\phi(y)=\sqrt{\frac{3b}{4}}\bigg[i\sqrt2\big[\text{E}(i\alpha y;1-72kb^2\alpha^2)\nonumber\\
 &&~~~~~~~~~-\text{F}(i\alpha y;1-72kb^2\alpha^2)\big]+\nonumber\\
  &&\sqrt{1+72kb^2\alpha^2+(1-72kb^2\alpha^2)\cosh(2\alpha y)}\tanh(\alpha y)\bigg],\nonumber\\
\end{eqnarray}
where $\text{F}(i\alpha y;1-72kb^2\alpha^2), \text{E}(i\alpha y;1-72kb^2\alpha^2)$ are the first and second kind elliptic integrals, respectively.  One can prove that $\phi(y)$ is real provided that $1-72kb^2\alpha^2\geq0$ as required by \eqref{constraint}. Specially, when $1-72kb^2\alpha^2=0$,
\begin{equation}\label{special}
    \phi(y)=\sqrt{\frac{3b}{2}}\tanh(\alpha y),
\end{equation}
which is a kink solution. However, for large enough $1-72kb^2\alpha^2$, the solution \eqref{solution2} turns to be a double-kink, as shown in Fig.\ref{phi}.

For other values $n$, an analytical solution like \eqref{solution2} is hard to obtain, but when
$\Delta=1$, \eqref{main}
reduces to
\begin{equation}
   \frac{3}{2}b\alpha^2\text{sech}^2(\alpha y)- \frac{3}{2}b\alpha^2\text{sech}^2(\alpha y)\tanh^{2n-2}(\alpha y)=\phi'^2.
\end{equation}
For $n=2$, it gives \eqref{special}.
For $n=3$,
\begin{eqnarray}
 &&\phi(y)=\sqrt{\frac{3b}{8}\cosh(2\alpha y)}\nonumber\\
 &&~~~~~~\times\bigg[2i\cosh^2(\frac{\alpha y}{2})\Xi \text{F}(i\text{arcsinh}(Z);17+12\sqrt{2})  \nonumber\\
 && +4i\cosh^2(\frac{\alpha y}{2})\Xi\Pi(3+2\sqrt{2};i\text{arcsinh}(Z),17+12\sqrt{2})\nonumber\\
 && ~~~~~~+\text{sech}(\alpha y)\tanh(\alpha y))\bigg],
\end{eqnarray}
where
\begin{eqnarray}
 &&Z\equiv\frac{\tanh(\frac{\alpha y}{2})}{\sqrt{3+2\sqrt{2}}},\nonumber\\
 && \Xi\equiv\text{sech}(2\alpha y)\sqrt{(3+2\sqrt{2})(1+Z^2)}\nonumber\\
 &&~~~~~\times\sqrt{1+(3+2\sqrt{2})^2Z^2},\nonumber
\end{eqnarray}
and $\Pi(3+2\sqrt{2};i\text{arcsinh}(Z),17+12\sqrt{2})$ is the third kind elliptical integral.
For $n=4$,
\begin{eqnarray}
  \phi(y) &&= i\sqrt{2b}\big[2\text{E}(2i\alpha y;\frac{3}{4})+\text{F}(2i\alpha y;\frac{3}{4})\big]\nonumber\\
  &&~~-\sqrt{b(5+3\cosh(\alpha y))}\tanh^3(\alpha y).
\end{eqnarray}

Generally, for different values $n$, via numerical approach, we find that the solution is a kink when $0<\Delta<1$ and a double kink as $\Delta$ is less than some negative value. From eq.~(\ref{main}), we can see that the domain wall solution turns out to be a double-kink solution when the contribution from the second term of right side exceeds the first term. So
$\Delta$ shows the strength of the geometrical effect of torsion.

\begin{figure}[h]
\includegraphics[width=7cm]{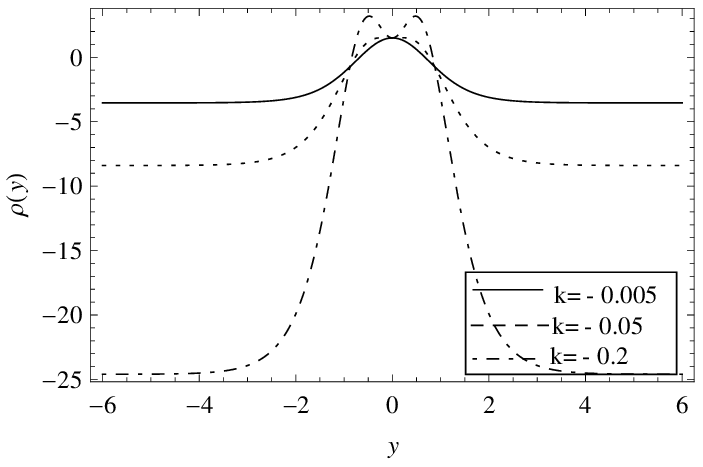}
\includegraphics[width=7cm]{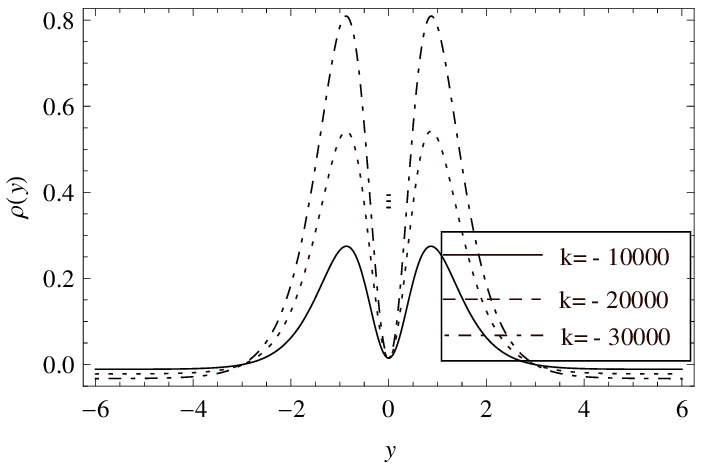}
 \caption{The density $\rho(y)$ of the scalar field $\phi(y)$ with $n=2, \alpha=1$, $b=1$ (top) and $b=0.01$ (bottom). A trend of brane-splitting can be seen from the transition.}\label{density}
\end{figure}

Commonly, the appearance of a double-kink solution means that the domain wall at $y=0$ symmetrically splits into two branes. This can be seen from the distribution of the energy density
\begin{eqnarray}
\rho(y)&&=\frac{3}{2}b\alpha^2\text{sech}^2(\alpha y)-3b^2\alpha^2\tanh^2(\alpha y)\nonumber\\
&&~~-(-3)^n2^{2n-3}kb^{2n-1}\alpha^{2n}(2n-1)\nonumber\\
&&~~\times(n\text{csch}^2(\alpha y)-2b)\tanh^{2n}(\alpha y),
\end{eqnarray}
as shown in fig.~\ref{density}. Locations of those two peaks are where two sub-branes inhabit. At the boundary of the spacetime
\begin{equation}
\rho(\pm\infty)=-3b^2\alpha^2+(-3)^n 2^{2n-2} k b^{2n} \alpha^{2n}(2n-1)
\end{equation}
is a minus constant if eq.~(\ref{constraint}) is satisfied.

\subsection{The split of brane}

In Ref.~\cite{Campos2002}, the authors investigated the split of thick brane, which is generated by a complex
scalar field coupled to gravity. They showed that the split of the brane is due
to a first-order transition when the temperature approaches the
critical value and a new disordered phase would appear between these two
sub-branes.

At zero temperature, the split of thick brane was realized by using a real scalar field \cite{BazeiaMenezesMenezes2003}. In this model, the engendered internal
structure depends on a real parameter, which changes the self-interaction of the scalar
field. The split of brane was also investigated by using two real scalar fields
~\cite{BazeiaFurtadoGomes2004,deSouzaDutra:2008gm,ZhaoLiu2010,Fu:2011pu,Liu:2011ysa}.

In our work, the internal structure is different, because the energy density of the scalar field is non-vanished at $y=0$, i.e., $\rho(0)=\frac{3}{2}b\alpha^2\neq 0$. Such a structure indicates that the split of the brane is incomplete, and there is a connection between the two sub-branes. While for the case with $\rho(0)=0$, the original brane is completely split, and the newly generated branes are independent. The energy density dwelling on the split branes becomes more notable with the increase of the contribution from torsion to which the phase transition is due. It indicates that the geometric effect will influence the distribution of the energy density. It should be noted that the distance between the two split branes is mainly determined by $\alpha$ and $b$, which also determine the thickness of the domain wall.

A probable explanation to the changeable $k$ is that $k$ might relates to the evolution of universe.
Note that the temperature of the cosmological background is a characteristic parameter relevant to the
evolution, so we can recognize $k$ as a function of temperature. Therefore, there might exists a critical temperature $T_c$, at which, the brane splits into two sub-branes.

\section{Localization of spin-$\frac{1}{2}$ particles}\label{section4}

 Whether various bulk fields could be confined to the brane by a
natural mechanism is an interesting and important issue to build up
the standard model. It has been known that massless scalar fields~\cite{Bajc2000} and gravitons~\cite{Randall1999,Gremm2000a,Csaki2000a} can be localized on
branes of different types. Abelian vector fields can be
localized on the RS brane in some higher-dimensional cases~\cite{Oda2000} or on thick dS branes and Weyl branes~\cite{Liu2008,Liu2008c}. The localization of fermion fields is also
interesting. In order to localize fermions, the coupling
between the fermion fields and the background scalars should be
introduced. With different scalar-fermion couplings, a single bound
state and a continuous gapless spectrum of massive fermion KK states
can be obtained, see for example~\cite{Liu2008a,Arias2002,Barbosa-Cendejas2005,Barbosa-Cendejas2006}.
In some other models, there exist finite discrete KK states
(mass gap) and a continuous gapless spectrum starting at a positive
$m^2$ \cite{Barbosa-Cendejas2008,Liu2009c,Liu2008c,Liu:2009ve,ZhaoLiu2011JHEP} or even only
exist bound KK modes.

From the results in Refs.~\cite{Liu2009c,Liu2008c}
we note that the effective potentials of the KK modes of scalar and
vector fields are free of gravity model and only dependent on $A(y)$.
It can be easily verified that the zero mode of these fields can be localized on the
brane we obtianed here. However, the effective potentials of fermion fields couple to the background scalar, so the localization is model-dependent.

In this section we will investigate how the spacetime torsion influences the localization of fermion fields on the brane. {Following the suggestion on fermion fields in Refs. \cite{Aldrovandi2007,Arcos:2005ec}, the equations will also be equivalent to the case in general relativity. Thus we can take the approach in general relativity, only in appropriate time we take our results into consideration.}
Via performing the conformal transformation $ dz=e^{-A(y)}dy$ \cite{Gremm2000a}, we can rerepresent the metric in conformal coordinates. Taking the simply Yukawa coupling, the 5-dimensional Dirac action of a massless
spin 1/2 fermion coupled to the background scalar $\phi$ is
\begin{equation}\label{fermionaction}
    S_{1/2}=\int{d^5x h (\bar{\Psi}\Gamma^M(\partial_M+\omega_M)\Psi-\eta\bar{\Psi}\phi\Psi)},
\end{equation}
The non-vanishing components of the spin connection $\omega_M$ for the background metric are
\begin{equation}
    \omega_{\mu}=\frac{1}{2}A'\gamma_\mu\gamma_5+\hat{\omega}_\mu,
\end{equation}
where prime denotes the derivation with respect to conformal coordinate $z$ from now on, and $\hat{\omega}_\mu$ is the spin connection on the brane and vanishes here.
Then the equation of motion is given by
\begin{equation}\label{4.3}
[\gamma^\mu \partial_\mu+\gamma^5(\partial_z+2A')-\eta e^A\phi]\Psi=0.
\end{equation}
The sign of the coupling $\eta$ of the spinor $\Psi$ to the scalar $\phi$ is arbitrary, and without loss of generality,
we assume $\eta>0$.

According to \eqref{4.3} $\Psi$ can be expanded by
\begin{equation}
\Psi=\sum_n[\Psi_{\textrm{L},n}(x)f_{\textrm{L},n}(z)+\Psi_{\textrm{R},n}(x)f_{\textrm{R},n}(z)]e^{-2A}
\end{equation}
with $\Psi_\textrm{L}=-\gamma^5\Psi_\textrm{L}$ and $\Psi_\textrm{R}=\gamma^5\Psi_\textrm{R}$ being the left-handed and right-handed components of a
4D Dirac field respectively. By demanding $\Psi_\textrm{L,R}$ satisfy the 4D massive Dirac equations $\gamma^\mu\partial_\mu\Psi_\textrm{L,R}=m\Psi_\textrm{L,R}$, we yield the following coupled equations
\begin{eqnarray}
&&[\partial_z+\eta e^A\phi]f_\textrm{L}(z)=mf_\textrm{R}(z),\\
&&{[\partial_z -\eta e^A\phi]f_\textrm{R}(z)=-mf_\textrm{L}(z)}.
\end{eqnarray}
These equations can be reduced to the Schr\"{o}dinger-like equations for the KK modes of left and
right chiral fermions
\begin{eqnarray}
&&[-\partial^2_z+V_\textrm{L}(z)]f_\textrm{L}(z)=m^2f_\textrm{L}(z),\\
&&[-\partial^2_z+V_\textrm{R}(z)]f_\textrm{R}(z)=m^2f_\textrm{R}(z),
\end{eqnarray}
where the effective potentials are given by
\begin{eqnarray}\label{potential}
&&V_\textrm{L}(z)=(\eta e^A\phi)^2-\eta\partial_z(e^A\phi),\\
&&V_\textrm{R}(z)=V_\textrm{L}(z)|_{\eta\rightarrow-\eta}.
\end{eqnarray}
The index $n$ is dropped for convenience.

\begin{figure}[h]
\begin{center}
\includegraphics[width=7.4cm]{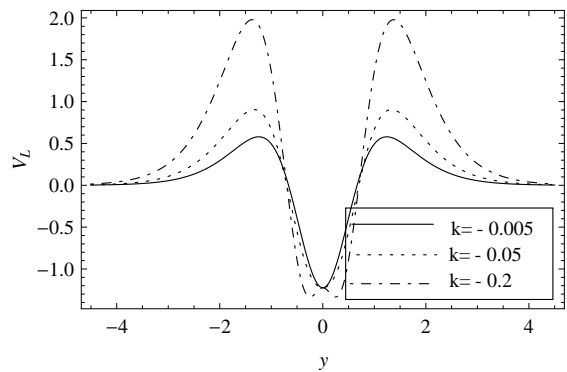}
\includegraphics[width=7.4cm]{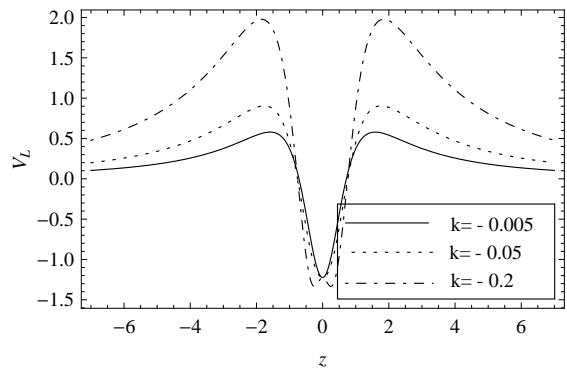}
\end{center}
 \caption{The effective potential of left-chiral fermions with $n=2, b=1, \alpha=1, {\eta=1}$ in different coordinate systems. }\label{localization}
\end{figure}

Since the Yukawa coupling is
an odd function of the extra dimension $z$, the
effective potential $V_\textrm{L,R}(z)$ of left- and right-chiral fermions
are invariant under the reflection
symmetry $z\rightarrow -z$. Here we discuss the case $n=2$, and get the effective potential
in proper coordinate $y$ from eqs.~(\ref{condition}) and (\ref{solution2}):
\begin{eqnarray}
 V_\textrm{L}(y)&=&\frac{1}{4}\cosh^{-2b-2}(\alpha y)
  \bigg[ \sqrt{3b}\alpha\eta
         \Big(2b\zeta\sinh^2(\alpha y)\nonumber\\
         &&+i\sqrt{2}b\xi\sinh(2\alpha y)-2\zeta\Big)\nonumber\\
     &&+3b\eta^2
         \Big(\zeta^2\sinh^2(\alpha y)+i\sqrt{2}\zeta\xi\sinh(2\alpha y)\nonumber\\
         &&-2\xi^2\cosh^2(\alpha y)\Big)
   \bigg],
\end{eqnarray}
where
\begin{eqnarray}
\xi&=&\text{E}(i\alpha y|1-72kb^2\alpha^2)-\text{F}(i\alpha y|1-72kb^2\alpha^2),\nonumber\\
\zeta&=&\sqrt{1+72kb^2\alpha^2+(1-72kb^2\alpha^2)\cosh(2\alpha y)}.\nonumber
\end{eqnarray}
For simplicity, we take $b=1$, and obtain the analytical transformation $y=\frac{\text{arcsinh}[z \alpha ]}{\alpha }$. Further, we can reexpress $V_\textrm{L}(y)$ as the function of the conformal coordinate $z$, $V_\textrm{L}(z)$ and $V_\textrm{L}(z)$ are plotted in Fig.~\ref{localization} for different values of $k$.

Since $b=1$, we can get a simple expression of the warped factor in conformal coordinate, i.e., $A(z)=-\ln\sqrt{1+z^2 \alpha ^2}$. Note that at $z=0$, $A(0)=A'(0)=\phi(0)=0$,
so $V_\textrm{L}(0)=-\eta e^{A(0)}\phi'(0)=-\sqrt{\frac{3}{2}}\eta\alpha$. For positive $\eta$, $V_\textrm{L}(0)<0$.
From fig.~\ref{localization}, it can be seen that $V_\textrm{L}(\pm\infty)$ vanishes at infinity, therefore there is only one bound massless mode for left-chiral fermions followed by a continuous gapless spectrum of KK states with $ \eta> 0$.

From fig.~\ref{localization}, we find that, with the decrease of $k$, the height of the potential well will
increase, then there will be two minima in the potential well, namely double well.
Although there is only one bound massless mode, but
some resonances may appear which can tunnel from the brane to
the bulk. In the case shown in fig. \ref{localization}, only the zero mode exists. But for {color{blue}$k=-0.5$} there exists an extra resonance with $m^2=3.2369$, probability $0.472493$ and odd wavefuction. For $k=-1.5$, there are two resonances with
$m^2=5.2925, 10.3974$. For $k=-4$, the resonances increase to three, $m^2=7.84658, 17.899, 25.032$. So the resonant states will increase with the contribution from torsion. For the right-chiral KK modes, there have no bound modes but continuous and gapless spectra which are the same with
the left-chiral KK modes.

Note that the height of the potential well will also increase with $b$ and $\alpha$, but
the width becomes narrower, then there will be a double well with a smaller $V_\textrm{L}(0)$.
The coupling constant $\eta$ can also affect the width and the height of the well, but there will be no transition
from one minimum in the potential well to two minima. Similarly the well has a smaller $V_\textrm{L}(0)$.

Next we discuss the condition of the localization. The zero mode for the left-chiral fermions reads~\cite{Liu:2009ve,ZhaoLiu2011JHEP,Fu:2011pu}
\begin{equation}\label{fermionsolution}
    f_{\textrm{L}0}(z)\propto \exp\bigg(-\eta\int_{0}^{z}{d\omega e^{A(\omega)}\phi(\omega)}\bigg).
\end{equation}
In order to check whether the zero mode can be localized on the brane, we should check
whether the normalization condition for the zero mode is satisfied, namely, whether the
integral
\begin{equation}\label{integral}
    \int{ f_{\textrm{L}0}(z)^2dz}\propto \int{\exp\bigg(-2\eta\int_{0}^{z}{d\omega e^{A(\omega)}\phi(\omega)}\bigg)dz}
\end{equation}
is finite. Since $e^{A(\omega)}\phi(\omega)\rightarrow 0$ when $\omega\rightarrow\infty$, so it is clear that the integral~(\ref{integral})
is finite for positive $\eta$, namely, the zero mode for left-chiral
fermions can be localized on the brane for positive $\eta$.

Since $b=1$, we can get a simple expression of the warped factor in conformal coordinate, i.e., $A(z)=-\ln\sqrt{1+z^2 \alpha ^2}$. At the infinity, $e^A\rightarrow\frac{1}{\alpha|z|}$, hereby,
\begin{equation}
    f_{\textrm{L}0}(z\rightarrow\pm\infty)\rightarrow|z|^{-\frac{\eta\phi_\infty}{\alpha}},
\end{equation}
where $\phi_\infty$ is
\begin{eqnarray}
    \phi(z\rightarrow\infty)\rightarrow\sqrt{\frac{3}{2}}\bigg[-i\text{E}(1-72k\alpha^2)+\text{K}(72k\alpha^2)\nonumber\\
    +\frac{i(1-72k\alpha^2)\text{E}(\frac{1}{1-72k\alpha^2})+i72k\alpha^2\text{K}(\frac{1}{1-72k\alpha^2})}{\sqrt{1-72k\alpha^2}}\bigg].
\end{eqnarray}
If the normalization condition is satisfied, we can get the following equivalent condition,
\begin{equation}\label{normalization}
    \int{|z|^{-\frac{2\eta\phi_\infty}{\alpha}}dz<\infty}.
\end{equation}
Only when $\eta>\eta_0=\frac{\alpha}{2\phi_\infty}$, the above
integral is convergent, which means that the left-chiral zero mode can be localized on the
brane under this condition.

From {eq.~(\ref{constraint})}, we can find that the consequences here can also be obtained for the even integer $n$.
For the odd integer $n$ and with $k$ bigger than some positive value, we also obtain the similar consequences here. Note that $k$ represents the strength of the contribution from torsion, so the results are applicable only when the torsion have a significant effect.

\section{Conclusion}

  In this paper, we investigate the geometric effect of torsion on thick branes in gauge theory, and find some analytic domain wall solutions for some specific values of $n$. We also find that the geometric effect determines whether the domain wall solution is a kink or double-kink. With the increase of the contribution of torsion, the configuration of the solution changes from a kink to double kink. The more significant the effect is, the more energy dwells on the sub-branes. We also study the
localization of fermion fields on the brane described by the domain wall solution. It is shown that there is only one bound massless mode on the brane, but when the spacetime torsion has a significant effect, the potential well for the fermion KK modes will become
more and more deeper and more resonant states of left-chiral
fermions with short lifetime will appear. With the coupling parameter $k$ being a function of temperature, the evolution of universe can influence not only the split behavior of the thick brane via changing the contribution from the spacetime torsion, but also the number of fermion resonate states.

\section*{Acknowledgments}

This work was supported by the National Natural Science Foundation of China (No. 11075065),
the Fundamental Research Funds for the Central Universities (No. lzujbky-2012-k30),
and the Natural Science Foundation of Gansu Province,
China (No. 096RJZA055).

\providecommand{\href}[2]{#2}\begingroup\raggedright
\endgroup


\end{document}